\documentclass[fleqn,usenatbib,onecolumn]{mnras}

\usepackage{newtxtext,newtxmath}

\usepackage[T1]{fontenc}
\usepackage{ae,aecompl}

\usepackage{graphicx}           % Including figure files.
\usepackage{amsmath}            % Advanced maths commands.
\usepackage{collcell}           % Included for placing diagrams in an array.
\usepackage{float}              % Used to disable figure floating environment when needed.
\usepackage{flafter}            % Floating elements after text
\usepackage{caption}            % Used for table footnotes.
\usepackage{booktabs}           % Supports threeparttable package below.
\usepackage{threeparttable}     % Supports adding white space to tables; used for improved spacing.

\newcommand{\dmunits}{\,pc\,cm$^{-3}$\:}    % Dispersion Measure units.
\newcommand{\funits}{\,Jy\,ms\:}            % Fluence units.
\newcommand{\DefaultPlotSize}{0.45}         % Default graphic scaling.
\newcommand{\BigPlotSize}{0.50}             % Default graphic scaling.
\newcommand{\LargestPlotSize}{0.65}         % Larger graphic size.
\newcommand{\uc}[1]{\,$\pm$\:#1}            % Macro to standardise formatting of uncertanties.
\newcommand{\uclu}[2]{\,$_{-#1}^{+#2}$\:}   % Two parameter uncertainty for different upper/lower values.
\newcommand{\FRB}[1]{FRB\,#1}               % Common formatting for a FRB.

\newcommand{\includepic}[1]{\includegraphics[width=6.0cm,height=5.0cm,keepaspectratio=true]{#1}}
\newcolumntype{i}{@{\hspace{1ex}}>{\collectcell\includepic}c<{\endcollectcell}}

\defcitealias{Macquart_Ekers_2018_2}{\,M\&E\:2018 }
\newcommand{\ME}{\citetalias{Macquart_Ekers_2018_2}}

\title[The FRB DM Distribution]{The Fast Radio Burst Dispersion Measure Distribution}

\author[Arcus \textit{et al.}]{
W. R. Arcus,$^{1}$\thanks{E-mail: wayne.arcus@icrar.org (WA)},
J.-P. Macquart$^{1}$, M. W. Sammons$^{1}$, C. W. James$^{1}$, R. D. Ekers$^{1,2}$
\\
$^{1}$ International Centre for Radio Astronomy Research, Curtin University, GPO Box U1987, Perth, WA 6845, Australia\\
$^{2}$ CSIRO Astronomy and Space Science, P.O. Box 76, Epping, NSW 1710, Australia\\
}

\date{Accepted XXX. Received YYY; in original form ZZZ}

\pubyear{2020}

\begin{document}
\label{firstpage}
\pagerange{\pageref{firstpage}--\pageref{lastpage}}
\maketitle

\begin{abstract}
We compare the dispersion measure (DM) statistics of FRBs detected by the ASKAP and Parkes radio telescopes. We jointly model their DM distributions, exploiting the fact that the telescopes have different survey fluence limits but likely sample the same underlying population. After accounting for the effects of instrumental temporal and spectral resolution of each sample, we find that a fit between the modelled and observed DM distribution, using identical population parameters, provides a good fit to both distributions. Assuming a one-to-one mapping between DM and redshift for an homogeneous intergalactic medium (IGM), we determine the best-fit parameters of the population spectral index, $\hat{\alpha}$, and the power-law index of the burst energy distribution, $\hat{\gamma}$, for different redshift evolutionary models. Whilst the overall best-fit model yields $\hat{\alpha}=2.2_{-1.0}^{+0.7}$ and $\hat{\gamma}=2.0_{-0.1}^{+0.3}$, for a strong redshift evolutionary model, when we admit the further constraint of $\alpha=1.5$ we favour the best fit $\hat{\gamma}=1.5 \pm 0.2$ and the case of no redshift evolution. Moreover, we find no evidence that the FRB population evolves faster than linearly with respect to the star formation rate over the DM (redshift) range for the sampled population.
\end{abstract}


%%%%%%%%%%%%%%%%% BODY OF PAPER %%%%%%%%%%%%%%%%%%
\section{Introduction}
Ever since their discovery by \cite{Lorimer_etal_2007}, it has been conjectured that the large dispersion measures (DMs) of fast radio bursts (FRBs) encode information about their distances and evolutionary history \citep[e.g., see the discussion in][]{Macquart_Ekers_2018_2}. Early suppositions that DMs contain a sizeable contribution due to their passage through the intergalactic medium (IGM), thereby placing the population at cosmological distances \citep{Lorimer_etal_2007, Thornton_etal_2013}, have been vindicated by recent localisations of two repeaters \citep{Chatterjeeetal2017,Marcote_etal_2020} and at least seven single events \citep[][]{Bannister_etal_2019,Prochaska_etal_2019,Ravi_etal_2019,Macquart_etal_2020}.

The cosmological nature of the FRB population has been further substantiated by the discovery of a relation between mean DM and fluence, $F$, in the bright non-repeating FRB population observed by the Australian SKA Pathfinder (ASKAP) and Parkes radio telescopes \citep{Shannon_etal_2018}. Therein, the authors note the average DM of the ASKAP sample, 440\,pc\,cm$^{-3}$, is half that of the fainter bursts detected by Parkes at 881\dmunits. This result is akin to the redshift-flux density relations observed in other cosmological populations -- for active galactic nuclei see the discussion in \citet{vonHoerner1973}. Further, \citet{vonHoerner1973} draws attention to the critical value of the luminosity function (in our case energy distribution) of $\gamma \approx 2.5$ when sources of a given fluence are equally probable in distance (in Euclidean space). The relevance of this will be discussed in Section 4.
% The existence of such a relation further challenges predictions that the host galaxies of FRBs make large (a few hundred $\text{pc cm}^{-3}$) contributions to the DM, whereby one would expect a burst DM distribution, dominated by the host galaxies, to be invariant to burst fluence thus contain little information on the burst distance.

The DM distribution is an observationally underexploited means of probing both the nature of FRB emission and the media through which they propagate. The redshift distribution of FRBs detected in a survey of finite sensitivity is shaped by the underlying burst luminosity, the evolution with redshift and by the spectral index of the emission. The DM distribution provides an additional means of accessing information on the mapping between redshift and DM, which is not bijective (i.e., one-to-one) except when averaged over many lines of sight \citep{Ioka_2003,Inoue_2004,McQuinn_2014}.

An especially powerful approach is to compare the DM distribution of two sample sets obtained from telescopes with significantly different detection thresholds. The DM-fluence relation noted by \cite{Shannon_etal_2018} exploits the first moment of the DM distribution. However, comparison of the shapes of the DM distributions obtained by surveys of differing sensitivities permits greater leverage to isolate key variables, since both distributions must be drawn from the same underlying luminosity and redshift distribution and with the same spectral index and host DM distributions.

It is known that the nature of the DM-fluence relation is a particularly useful probe of the average burst luminosity distribution\footnote{The nature of the relation also depends upon the parameters of the cosmological model \citep{vonHoerner1973}, however these are not regarded as free parameters in the present treatment.} \citep{Macquart_Ekers_2018_2}. Whether the relation manifests as a correlation or anti-correlation depends upon the slope of the burst luminosity function, while the scatter of bursts about the relation contains information on the intrinsic spread of burst luminosities \citep{Lorimer_etal_2007,Shannon_etal_2018}.

It is the purpose of this paper to examine the DM distributions of the FRB populations detected by the ASKAP and Parkes radio telescopes. In Section 2, we present the samples used in our analysis, the DM histograms of those sample sets, a summary of the formalism relating the observed distributions to survey parameters and the underlying properties of the FRB distribution. In Section 3, we employ this formalism to infer the properties of the FRB distribution, the mapping between DM and distance and the effect of instrument performance on the detectability of bursts as a function of DM. The implications of our results and conclusions are discussed in Sections 4 and 5 respectively.

%==================================
\section{FRB Dispersion Measure Distributions}
The treatment herein is based on the analysis of FRB event data detected by the Commensal Real-time ASKAP Fast Transients (CRAFT) survey and by various surveys with the Parkes radio telescope. The ASKAP-CRAFT data are drawn from \citet{Shannon_etal_2018} and \cite{Macquart_etal_2019}, whilst Parkes data are drawn from FRBCAT \citep[][\url{http://frbcat.org}]{Petroff_etal_2016} and are summarised in Table \ref{tab:FRBdata}. In our analysis, we exclude the Lorimer Burst (\FRB{010724}) to avoid potential discovery bias, as discussed in \citet{Macquart_Ekers_2018_1}, although it is unclear as to the extent to which such a bias may affect the DM. (We include FRBs that are below the nominal fluence limits for their respective telescopes, since these limits are characteristic values averaged over telescope parameters -- in particular, beam-shape.) We utilise the DM of the IGM, $DM_{\text{IGM}}$, determined via eq.(\ref{eqn:CorrectedDM}), by estimating and removing the DM contributions due to the Milky Way disc, $DM_{\text{MW}}$, its halo, $DM_{\text{Halo}}$, and the FRB Host environment, $DM_{\text{Host}}$, thus:

\begin{equation}
DM_{\text{Obs}} = DM_{\text{MW}} + DM_{\text{Halo}} + DM_{\text{IGM}} + DM_{\text{Host}}/(1+z)\text{.}
\label{eqn:CorrectedDM}
\end{equation}
\\
For the ASKAP-CRAFT (lat50 survey) data, we assume $DM_{\text{MW}}\approx30$ \dmunits, due to the high galactic latitudes of the observations \citep[NE2001,][]{Cordes_Lazio_2003}, $DM_{\text{Halo}} \approx 30$ \dmunits and $DM_{\text{Host}} \approx 50$ \dmunits throughout \citep{Dolag_etal_2015,Xu_Han_2015,Tendulkar_etal_2017,Mahony_etal_2018,Macquart_etal_2020}. We note the assumption that $DM_{\text{Host}} \approx 50$ \dmunits, instead of using a distribution of possible $DM_{\text{Host}}$ values, will only have a small effect given the much larger observed DM values utilised in this analysis. %Moreover, the notion that the contribution of the $DM_{\text{Host}}$ is relatively small component of the overall DM is supported by the DM-fluence relation noted by \citet{Shannon_etal_2018}.
Values of $DM_{\text{MW}}$ for individual Parkes events are drawn from FRBCAT.

\begin{table*}
%\centering
\caption{Summary of the Parkes and ASKAP-CRAFT FRB events extracted from \citet{Shannon_etal_2018}, \citet{Macquart_etal_2019} and FRBCAT \citep[]{Petroff_etal_2016} utilised herein. Survey References for individual FRBs are:
(1) \citet{Spolaor_Bannister_2014}; (2) \citet{Zhang_etal_2019}; (3) \citet{Keane_etal_2011}; (4) \citet{Champion_etal_2016}; (5) \citet{Petroff_etal_2016}; (6) \citet{Thornton_etal_2013}; (7) \citet{Ravi_etal_2015}; (8) \citet{Petroff_etal_2015}; (9) \citet{Petroff_etal_2016};(10) \citet{Keane_etal_2016}; (11) \citet{Bhandari_etal_2018}; (12) \citet{Ravi_etal_2016}; (12) \citet{Shannon_etal_2018}; (13) \citet{Oslowski_etal_2019}; \& (14) \citet{Macquart_etal_2019}. Note: (i) fluences of the Parkes events are lower limits; (ii) the Lorimer Burst (\FRB{010724}) has been excluded from our analysis to avoid potential discovery bias; and (iii) we assume $\text{DM}_{\text{MW}} \approx 30$ \dmunits for the ASKAP FRBs of \citet{Shannon_etal_2018} due to the high galactic latitude of the lat50 ASKAP-CRAFT survey.}
\begin{tabular}{lccccc}
\hline
Designation & $\text{DM}$ & $\text{DM}_{\text{MW}}$ & Fluence & Survey & Telescope \\ & (pc\,cm$^{-3}$) & (pc\,cm$^{-3}$) & (Jy\,ms) & Reference \\
\hline
\FRB{010125}    & 790   & 110   & $>2.82$   & 1    & Parkes    \\
\FRB{010312}    & 1187  & 51    & $>6.1$    & 2    & Parkes    \\
\FRB{010621}    & 745   & 523   & $>2.87$   & 3    & Parkes    \\
\FRB{090625}    & 900   & 32    & $>2.18$   & 4    & Parkes    \\
\FRB{110214}    & 169   & 31    & $>51.3$   & 5    & Parkes    \\
\FRB{110220}    & 944   & 35    & $>7.28$   & 6    & Parkes    \\
\FRB{110626}    & 723   & 47    & $>0.89$   & 6    & Parkes    \\
\FRB{110703}    & 1104  & 32    & $>2.15$   & 6    & Parkes    \\
\FRB{120127}    & 553   & 32    & $>0.55$   & 6    & Parkes    \\
\FRB{121002}    & 1629  & 74    & $>2.34$   & 4    & Parkes    \\
\FRB{130626}    & 952   & 67    & $>1.47$   & 4    & Parkes    \\
\FRB{130628}    & 470   & 53    & $>1.22$   & 4    & Parkes    \\
\FRB{130729}    & 861   & 31    & $>3.43$   & 4    & Parkes    \\
\FRB{131104}    & 779   & 71    & $>2.33$   & 7    & Parkes    \\
\FRB{140514}    & 563   & 35    & $>1.32$   & 8    & Parkes    \\
\FRB{150215}    & 1106  & 427   & $>2.02$   & 9    & Parkes    \\
\FRB{150418}    & 776   & 189   & $>1.76$   & 10    & Parkes    \\
\FRB{150610}    & 1594  & 122   & $>1.3$    & 11    & Parkes    \\
\FRB{150807}    & 266   & 37    & $>44.8$   & 12    & Parkes    \\
\FRB{151206}    & 1910  & 160   & $>0.9$    & 11    & Parkes    \\
\FRB{151230}    & 960   & 38    & $>1.9$    & 11    & Parkes    \\
\FRB{160102}    & 2596  & 13    & $>1.8$    & 11    & Parkes    \\
\FRB{170107}    & 610   & 30    & 58        & 12    & ASKAP     \\
\FRB{170416}    & 523   & 30    & 97        & 12    & ASKAP     \\
\FRB{170428}    & 992   & 30    & 34        & 12    & ASKAP     \\
\FRB{170707}    & 235   & 30    & 52        & 12    & ASKAP     \\
\FRB{170712}    & 313   & 30    & 53        & 12    & ASKAP     \\
\FRB{170906}    & 390   & 30    & 74        & 12    & ASKAP     \\
\FRB{171003}    & 463   & 30    & 81        & 12    & ASKAP     \\
\FRB{171004}    & 304   & 30    & 44        & 12    & ASKAP     \\
\FRB{171019}    & 461   & 30    & 219       & 12    & ASKAP     \\
\FRB{171020}    & 114   & 30    & 200       & 12    & ASKAP     \\
\FRB{171116}    & 618   & 30    & 63        & 12    & ASKAP     \\
\FRB{171209}    & 1458  & 13    & $>2.3$    & 13    & Parkes    \\
\FRB{171213}    & 159   & 30    & 118       & 12    & ASKAP     \\
\FRB{171216}    & 203   & 30    & 36        & 12    & ASKAP     \\
\FRB{180110}    & 716   & 30    & 380       & 12    & ASKAP     \\
\FRB{180119}    & 403   & 30    & 100       & 12    & ASKAP     \\
\FRB{180120.2}  & 496   & 30    & 60        & 12    & ASKAP     \\
\FRB{180120}    & 441   & 30    & 51        & 12    & ASKAP     \\
\FRB{180130}    & 344   & 30    & 104       & 12    & ASKAP     \\
\FRB{180131}    & 658   & 30    & 114       & 12    & ASKAP     \\
\FRB{180212}    & 168   & 30    & 108       & 12    & ASKAP     \\
\FRB{180309}    & 263   & 45    & $>12$     & 13    & Parkes     \\
\FRB{180311}    & 1576  & 45    & $>2.4$    & 13    & Parkes    \\
\FRB{180315}    & 479   & 116   & 11        & 14    & ASKAP     \\
\FRB{180324}    & 431   & 70    & 71        & 14    & ASKAP     \\
\FRB{180714}    & 1470  & 257   & $>5$      & 13    & Parkes    \\
\hline
\end{tabular} \label{tab:FRBdata}
\end{table*}

The survey fluence limit at $DM=0$, $F_{0}$, of the Parkes and ASKAP telescopes are in the approximate ranges of 1-5 \funits and 21-31\funits respectively and are dependent upon the slope of the source counts distribution at their limits \citep[see Table 2 of][]{James_etal_2018}. While the DMs of the bursts are well determined for both FRB event datasets, the fluences of the Parkes events are lower limits due to the inherent inability to localise each burst within individual beams \citep{KeanePetroff2015,Macquart_Ekers_2018_1}. The Parkes fluences are therefore referenced to the beam centre. In practice, this limitation is not expected to significantly affect the present analysis since the modelled DM distributions are referenced only to a limiting survey depth and do not require information pertaining to each burst.

\begin{figure}
    \centering
    \includegraphics[scale=\DefaultPlotSize]{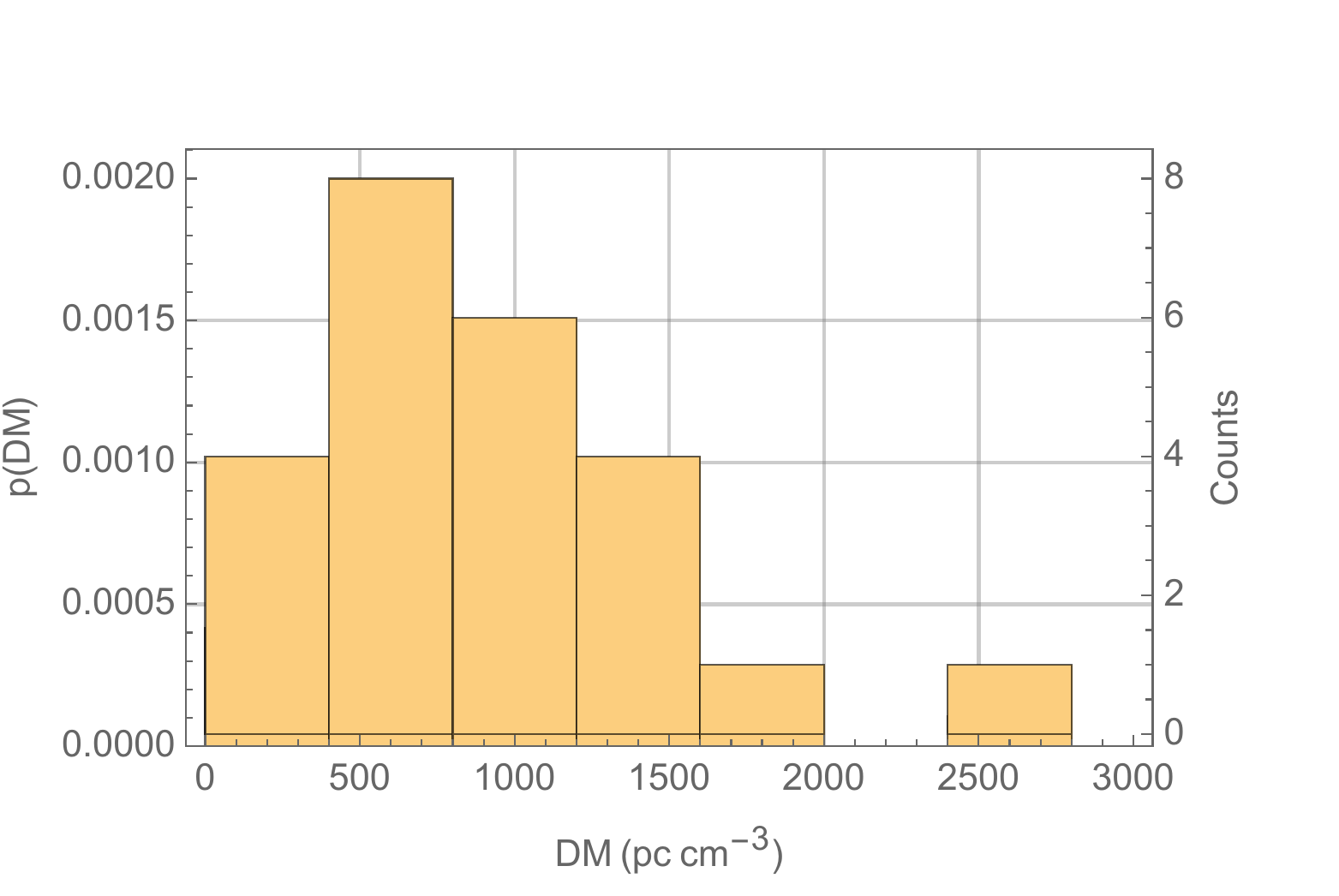}
    \includegraphics[scale=\DefaultPlotSize]{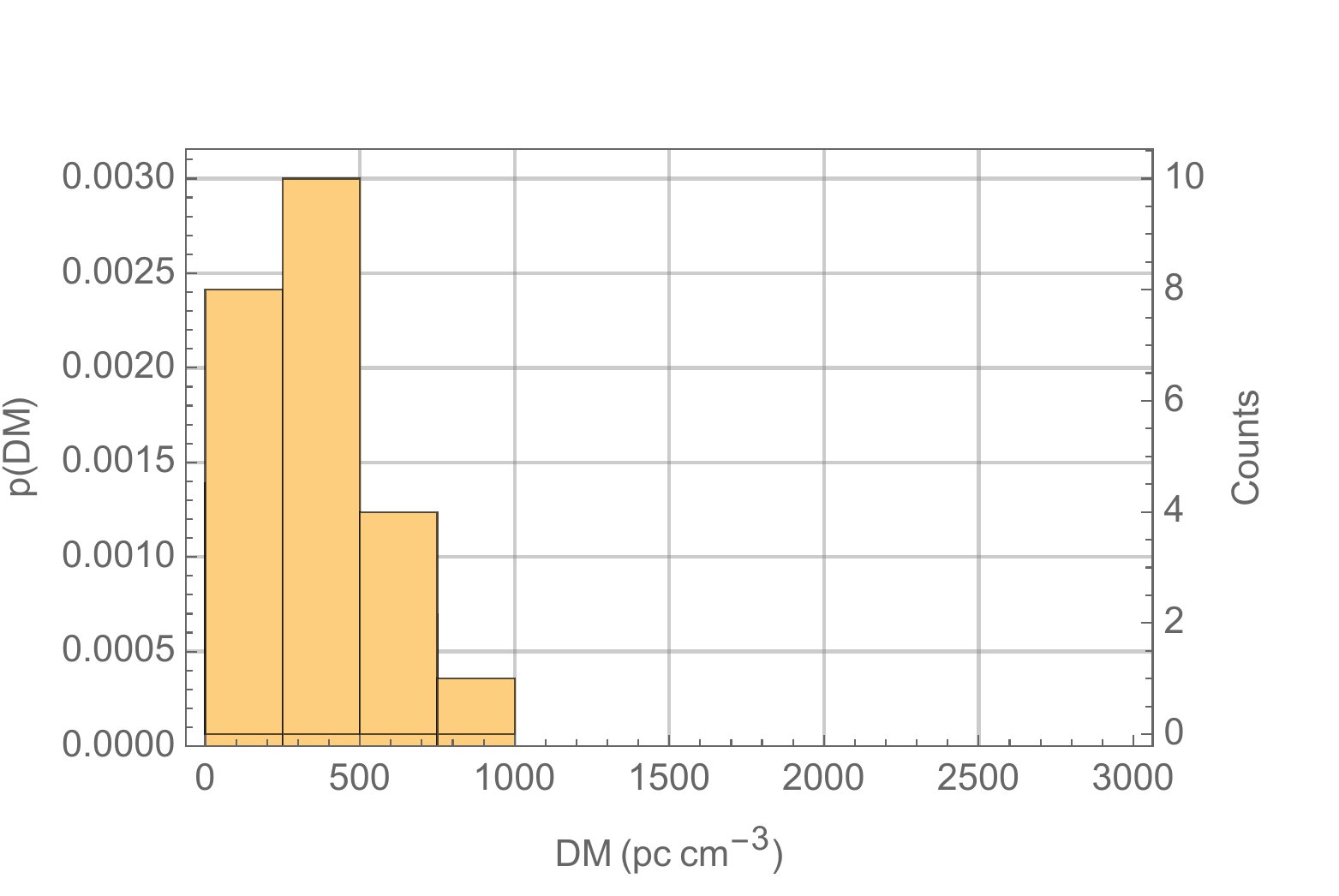}
    \caption{The DM histograms of the Parkes (left panel) and ASKAP (right panel) FRB events using a DM bin size of 300 \dmunits.}
    \label{fig:ParkesASKAPHistogram}
\end{figure}

Figure \ref{fig:ParkesASKAPHistogram} depicts the DM histograms of the Parkes and ASKAP FRB event samples listed in Table \ref{tab:FRBdata}. An interesting feature of the histograms is that they have similar \textit{shapes}. The means of these distributions differ by 594\:pc\,cm$^{-3}$, an update to the value of \cite{Shannon_etal_2018}, due to the increased sample size. Whilst these overall shapes are expected\footnote{The initial increase at low DM is due to the volume sampled increasing as distance cubed. The counts then decrease at higher DMs (distances) as the fluences of the less luminous bursts drop below the survey sensitivity limit -- i.e., they become incomplete.}, a quantitative analysis of the data necessitates we account for the finite instrumental spectral and temporal resolution of both telescope back-ends.

%==================================
\subsection{DM Distribution Model}
We utilise the model of \citet[][hereinafter the \ME model]{Macquart_Ekers_2018_2} by adopting the fluence-based formalism of eq.(\ref{eqn:DMDistribution}), and their symbols as defined in Table \ref{tab:ListOfSymbols}, to estimate the semi-constrained parameters of fluence spectral index, $\alpha$, energy power-law index, $\gamma$, and survey fluence limit at $DM=0$, $F_{0}$, for an assumed energy power-law regime. We compare the Parkes and ASKAP DM histograms with corresponding modelled DM distributions, $dR_{F}/dDM$, where $R_{F}$ is the total differential (fluence) event rate in the observer's frame ($dR_{F}/dDM$ has units of $\text{events s}^{-1} \; (\text{pc cm}^{-3})^{-1} \; \text{sr}^{-1} $).

The DM distribution, for fluences above a minimum survey fluence limit, is dependent upon a number of factors including the underlying redshift distribution of the population (relating to the evolutionary history) and the mean DM gradient, $d\overline{DM}/dz$, for an assumed homogeneous IGM. These are dependent upon the source energy distribution function, characterised by the source minimum and maximum energies $E_{\text{min}}$ and $E_{\text{max}}$ respectively, the energy power-law index and the spectral index \citep{Macquart_Ekers_2018_2}.

Throughout this work we utilise a $\Lambda$CDM universe with cosmological parameters and equations consistent with the \citet{Plank_Results_2013}\footnote{$(h,\: H_{0},\: \Omega_b,\: \Omega_m,\: \Omega_{\Lambda},\: \Omega_k) = (0.7,\: 100\,\text{h km s}^{-1} \text{Mpc}^{-1},\: 0.049,\: 0.318,\: 0.682,\: 0)$.} and \citet{Hogg_2000}, and symbol definitions of the \ME model with some minor notational changes and extensions as summarised in Table \ref{tab:ListOfSymbols}.

\begin{table*}
\caption{Symbol definitions relevant to the \ME model utilised herein with some minor notational changes and extensions.}
\begin{tabular}{ll}
\hline
Symbol                     & Definition \\
\hline
$G$                        & Gravitational constant \\
$m_{p}$                    & Proton rest mass \\
$z$                        & Redshift \\
$c$                        & Speed of light \textit{in vacuo} \\
$H_{0}$                    & Hubble constant at the present epoch \\
$E(z)$                     & Dimensionless Hubble parameter $E(z)=\sqrt{\Omega_{m} (1+z)^{3} + \Omega_{k} (1+z)^{2} + \Omega_{\Lambda}}$ \\
$H(z)$                     & Hubble constant at an arbitrary redshift $z : H(z) =H_{0} E(z)$ \\
$D_{H}$                    & Hubble distance \\
$D_{M}$                    & Comoving distance \\
$D_{L}$                    & Luminosity distance \\
$R_{F}$                    & Total (fluence) differential FRB event rate in the observer's frame \\
$\Omega_m$                 & Matter density (baryonic and dark) \\
$\Omega_{\Lambda}$         & Vacuum density \\
$\Omega_k$                 & Spatial curvature density \\
$\Omega_b$                 & Baryonic matter density \\
$\alpha$                   & Fluence spectral index defined such that $F_{\nu} \propto \nu^{-\alpha}$ \\
$\gamma$                   & Energy power-law index \\
$F_{0}$                    & Fluence survey limit at $DM=0$\\
$F_{\text{0,P}}$           & Fluence survey limit of the Parkes telescope at $DM=0$\\
$F_{\text{0,A}}$           & Fluence survey limit of the ASKAP telescope at $DM=0$\\
$F_{\nu}$                  & Fluence (energy spectral density per unit area)\\
$F_{\text{min}}$           & Minimum fluence for luminosity curve \\
$F_{\text{max}}$           & Maximum fluence for luminosity curve \\
$E_{\nu}$                  & Spectral energy density \\
$E_{\text{min}}$           & Lower spectral energy density bound for the event rate energy function \\
$E_{\text{max}}$           & Upper spectral energy density bound for the event rate energy function \\
$\displaystyle dR_{F}/dz$  & Fluence-based redshift distribution \\
$\displaystyle dR_{F}/dDM$ & Fluence-based DM distribution \\
$\overline{DM}(z)$         & Mean DM for the homogeneous IGM \\
$X_{\text{e,H}}$           & Fraction of ionised Hydrogen in the homogeneous IGM \\
$X_{\text{e,He}}$          & Fraction of ionised Helium in the homogeneous IGM \\
$\psi_{n}(z)$              & Event rate per comoving volume as a function of redshift: $\psi_{n}(z) \propto \Psi^{n}(z)$ \\
$\Psi(z)$                  & The cosmic star formation rate (CSFR) per comoving volume \\
$n$                        & Exponent of the redshift evolutionary term per comoving volume \\
\hline
\end{tabular}
\label{tab:ListOfSymbols}
\end{table*}

The FRB redshift and DM distributions, in a survey of limiting fluence, are respectively given by eq.(\ref{eqn:RedshiftDistribution}) \& eq.(\ref{eqn:DMDistribution})

\begin{equation}
    \dfrac{dR_{F}}{dz}(F_{\nu} > F_{0},z;\alpha,\gamma,n,F_{0},F_{\text{min}},F_{\text{max}}) = 4 \pi D_H^5\left(\dfrac{D_M}{D_H}\right)^4 \dfrac{(1+z)^{\alpha-1}}{E(z)} \psi_{n}(z)
    \begin{cases}
        0 & F_{0} > F_{\text{max}} \\
        \dfrac{(1+z)^{2-\alpha}}{4 \pi D_L^2}\left(\dfrac{F_{\text{max}}^{1-\gamma} - F_0^{1-\gamma}}{F_{\text{max}}^{1-\gamma} - F_{\text{min}}^{1-\gamma}}\right) & F_{\text{min}} \le F_{0} \le F_{\text{max}} \\
        \dfrac{(1+z)^{2-\alpha}}{4 \pi D_L^2} & F_{0} < F_{\text{min}}
    \end{cases}
    \label{eqn:RedshiftDistribution}
\end{equation}
\\

\noindent and

\begin{equation}
    \dfrac{dR_F}{dDM}(DM;\alpha,\gamma,n,F_{0},F_{\text{min}},F_{\text{max}}) = \dfrac{dR_F}{dz}(F_{\nu} > F_0, z;\alpha,\gamma,n,F_{0},F_{\text{min}},F_{\text{max}})/\dfrac{d\overline{DM}}{dz} \text{,}
\label{eqn:DMDistribution}
\end{equation}
\\

\noindent where the minimum and maximum of the source energy is related, respectively, to the corresponding minimum and maximum fluence via $E_{[\text{min/max}]} = 4 \pi D_{L}^{2}(z) F_{[\text{min/max}]}/((1+z)^{2-\alpha})$. Here, $\psi_{n}(z)$ and $\overline{DM}(z)$ represent redshift evolution \citep{Madau_Dickinson_2014}, via eq.(\ref{eqn:CSFR}), and the mean DM of an homogeneous IGM \citep{Inoue_2004,Ioka_2003}, via eq.(\ref{eqn:DM}), respectively

\begin{equation}
    \psi_{n}(z) = K \left(\dfrac{0.015 (1 + z)^{2.7}}{1 + ((1 + z)/2.9)^{5.6}}\right)^{n} \mathrm{yr^{-1} Mpc^{-3}}
    \label{eqn:CSFR}
\end{equation} \\

\noindent and

\begin{equation}
    \overline{DM}(z)=\dfrac{3H_0c\Omega_b}{8 \pi G m_p} \int_{0}^{z} \dfrac{(1+z') \left[\frac{3}{4} X_{\text{e,H}}(z') + \frac{1}{8}X_{\text{e,He}}(z')\right] }{\sqrt{(1+z')^3 \Omega_m + \Omega_{\Lambda}}} dz'\text{.}
    \label{eqn:DM}
\end{equation}
\\

Whilst the underlying redshift distribution of the FRB population is unknown, we follow \citet{Macquart_Ekers_2018_2} and adopt the \citet{Madau_Dickinson_2014} formalism for the cosmic star formation history of the Universe. Equation (\ref{eqn:CSFR}) accounts for the redshift evolution of the rate density for a progenitor population abundance, governed by stellar processes throughout cosmic history, via the relation $\psi_{n}(z) \propto \Psi^{n}(z)$. Here, $\Psi$ represents the cosmic star formation rate (CSFR) per comoving volume and the event rate per comoving volume, $\psi_{n}(z)$, is related via a power-law index, $n$ \citep{Macquart_Ekers_2018_2}.

We specifically consider three cases: (i) $\psi_{0}(z)$ -- no redshift evolution; (ii) $\psi_{1}(z)$ -- redshift evolution being linearly proportional to the CSFR; and (iii) $\psi_{2}(z)$ -- redshift evolution being quadratically proportional to the CSFR. The case of $\psi_{0}(z) = 1$ represents a constant event rate per comoving volume and $\psi_{2}(z)$ represents a rapidly evolving population. Furthermore, throughout this work, we set the ionised fraction of Hydrogen and Helium to $X_{\text{e,H}}=1$ for $z<8$ and $X_{\text{e,He}}=1$ for $z<2.5$ respectively, and zero otherwise, and take $\alpha$ to refer to the spectral index of the burst fluence unless specifically noted otherwise. The dimensionless Hubble parameter, $E(z)$, permits the Hubble parameter for an arbitrary redshift, $H(z)$, to be determined given the Hubble constant at the present epoch, $H_{0} : H(z) = H_{0} E(z)$.

%==================================
\section{DM Distribution Properties}
\subsection{Instrument Response}
The nominal fluence thresholds utilised for the CRAFT lat50 survey with ASKAP \citep{Shannon_etal_2018} and the SUPERB survey with the Parkes multibeam \citep{Keane_etal_2018} are $26$\funits and $2$\funits respectively, quoted for bursts of pulse-widths, $w$, 1.266 ms and 1.0 ms respectively. For bursts of a different width, the detection sensitivity varies as $w^{-1/2}$ due to extra noise (time) over which the burst energy is spread.

Here, we are principally concerned with DM-dependent effects, introduced by the different spectral and temporal resolutions used for incoherent de-dispersion searches in Parkes and ASKAP FRB surveys. To evaluate this effect, artificial bursts with a synthetic flat time-frequency profile were injected at random times into the time-frequency dynamic spectrum with temporal resolution, $t_{r}$, and spectral resolution, $\nu_{r}$, as given in Table \ref{tab:SensitivityFit} of Appendix A, and an incoherent de-dispersion performed. The sensitivity, $\eta$, of the telescope responses were parametrised via eq.(\ref{eqn:Sensitivity}), by comparing the recovered signal-to-noise ratio (SNR) with that expected from the radiometer equation,

\begin{equation}
    {\rm SNR_{\rm rad}} =  \frac{F_\nu}{\rm SEFD} \sqrt{\frac{2 \Delta \nu}{w} },
\end{equation}\\

\noindent for a system with spectral equivalent flux density (SEFD) and bandwidth, $\Delta \nu$, detecting a burst with width, $w$, and fluence, $F_\nu$. Converting to efficiency yields:

\begin{equation}
    \eta(DM,w) \equiv \frac{\rm SNR}{\rm SNR_{\rm rad}}
    \approx \dfrac{\eta_{0}}{ \sqrt{c_{1} 2 k \,\text{DM}\, \nu_{r}  \bar{\nu}^{-3}_{c} + c_{2} t_{r} + w }}\text{,}
\label{eqn:Sensitivity}
\end{equation}\\

\noindent where $\eta_{0}$, $c_{1}$ and $c_{2}$ are fitting constants for each telescope, $\bar{\nu}_{c}$, the dispersion-weighted mean frequency, and $k$ a constant relating time-delay to DM (see Table \ref{tab:SensitivityFit}). Here, $k=4.149$\,ms, where the time delay (ms) of a burst at frequency $\nu$ (GHz) is given by $\Delta t = k \text{DM} \nu^{-2}$ for a given DM (pc\,cm$^{-3}$).  We note that this formulation has been shown to reproduce the telescope performance \citep[see Supplementary material in][]{Shannon_etal_2018}.

The three terms in the denominator of eq.(\ref{eqn:Sensitivity}) represent, respectively, the smearing of burst fluence within a frequency channel due to its dispersion, the time resolution of the instrument and the intrinsic burst-width. The form is similar (but not identical) to the geometric addition of smearing terms used by \citet{Cordes_McLaughlin_2003} and subsequently found in much of the FRB literature. The mean sensitivity, $\bar{\eta}$, to a distribution of burst widths may be calculated by averaging $\eta$ over the distribution. We assume a log-normal distribution\footnote{The value of the sensitivity at $\text{DM}=0$ is unimportant for present purposes since the pulse-width is smaller than the instrument resolution and we are assessing the relative rates and the influence of curve shapes. The power of this approach means that this technique is insensitive to many of the specifics of the particular distribution chosen.} in $w$, producing:

\begin{equation}
    \bar{\eta}({DM}) =
        \dfrac{1}{\sqrt{2 \pi} \ln \sigma}
	       \int_{0}^{w_{m}}
		      \frac{1}{w} \eta({\text{DM}},w) e^{- (\ln w - \ln \mu)^{2} / (2 \ln \sigma)} dw,
\label{eqn:DMEfficiency}
\end{equation}
\\

\noindent where the maximum burst search width, $w_m$, is taken to be $32\,\text{ms}$.

The mean and standard deviations of the burst-width distribution (viz., $\mu=2.67$\,ms and $\sigma = 2.07$\,ms) were derived by simultaneously fitting the observed burst-width distribution of ASKAP and Parkes FRBs \citep{Petroff_etal_2016} in accounting for the finite resolution of the instruments \citep{Connor_2019}.

It is important to realise that a complete treatment of the sensitivity of a FRB search must incorporate the search efficiency, which is a function of DM, hence {\it the limiting fluence of any survey is also a function of DM}. Its effect is incorporated by using eq.(\ref{eqn:DMEfficiency}) in the survey fluence limit of equation eq.(\ref{eqn:RedshiftDistribution}), by mapping DM to redshift and making the substitution $F_{0} \rightarrow F'_{0}(z)$. In this approach, we utilise the relation $F'_{0}(z) = \{F_{0} / \overline{\eta}(z) : \overline{\eta}(z=0)=1\}$, where we normalise $\overline{\eta}(z)$ and interpret $F_{0}$ as $F'_{0}(z=0)$.

\begin{figure}
    \centering
    \includegraphics[scale=\LargestPlotSize]{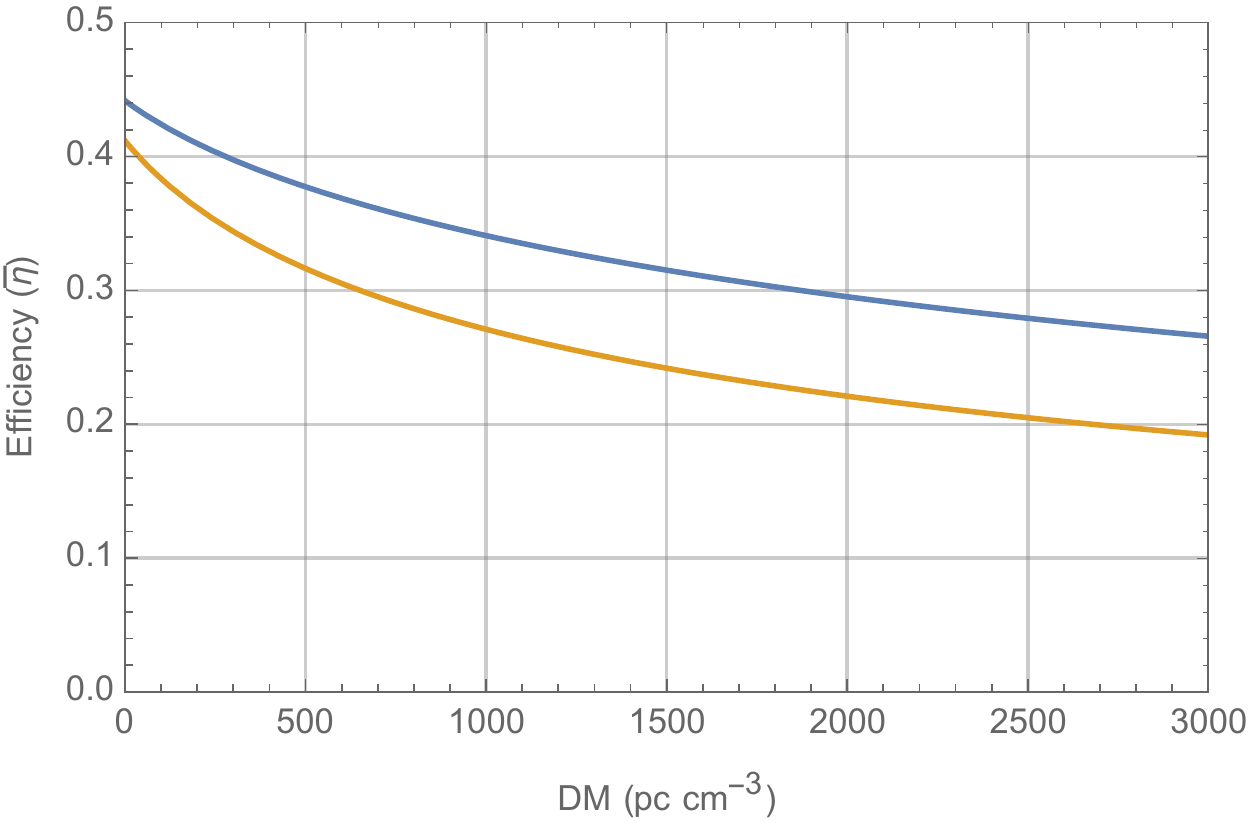}
    \caption{DM response curves for the Parkes (blue) and ASKAP (orange) telescopes using eq.(\ref{eqn:DMEfficiency}) with the parameters of Table \ref{tab:SensitivityFit} and for the population pulse-width mean $\mu=3.44$\,ms and standard deviation $\sigma=2.66$\,ms.}
    \label{fig:ResponseCurves}
\end{figure}

Figure \ref{fig:ResponseCurves} depicts the resultant DM response curves utilised for each telescope based on an assessment of the statistics for FRB events listed in Table \ref{tab:FRBdata}. Here, we are primarily interested in the relative efficiency between the Parkes and ASKAP telescopes and not the absolute FRB detection rates. As noted in \textsection3.2, we also check the robustness of the fitting process to pulse-widths less than the receiver time resolution by conducting the same parameter best-fit estimation using response curves for a mean pulse-width one decade lower than that of the FRB samples (i.e., $\mu = 0.334$\,ms). We observe that this is a high-order effect, resulting in the response curves shifting vertically whilst maintaining their overall shape -- an effect that is subsequently normalised out in the fitting process.

\begin{figure}
    \centering
    \includegraphics[scale=\BigPlotSize]{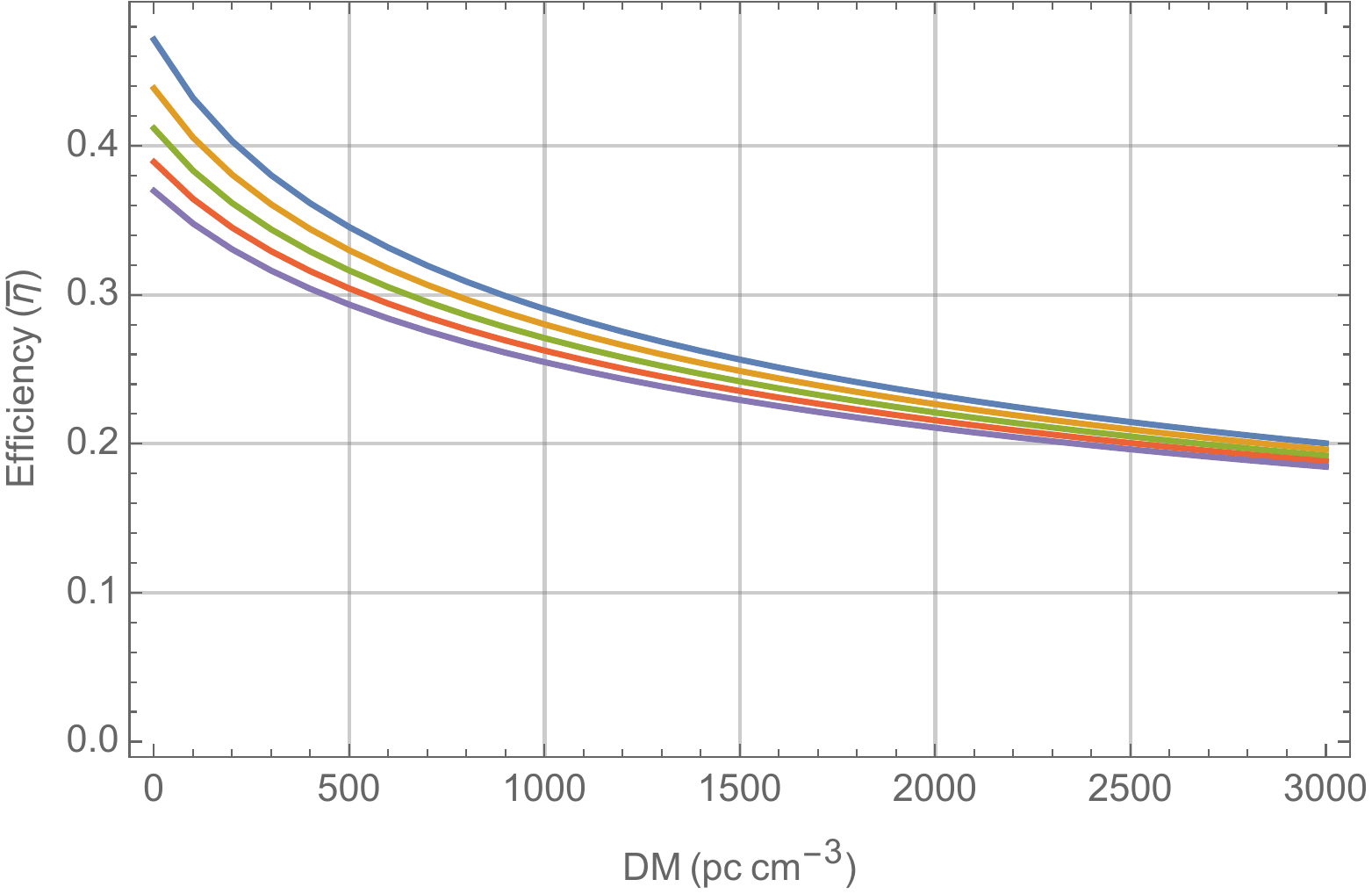}
    \includegraphics[scale=\BigPlotSize]{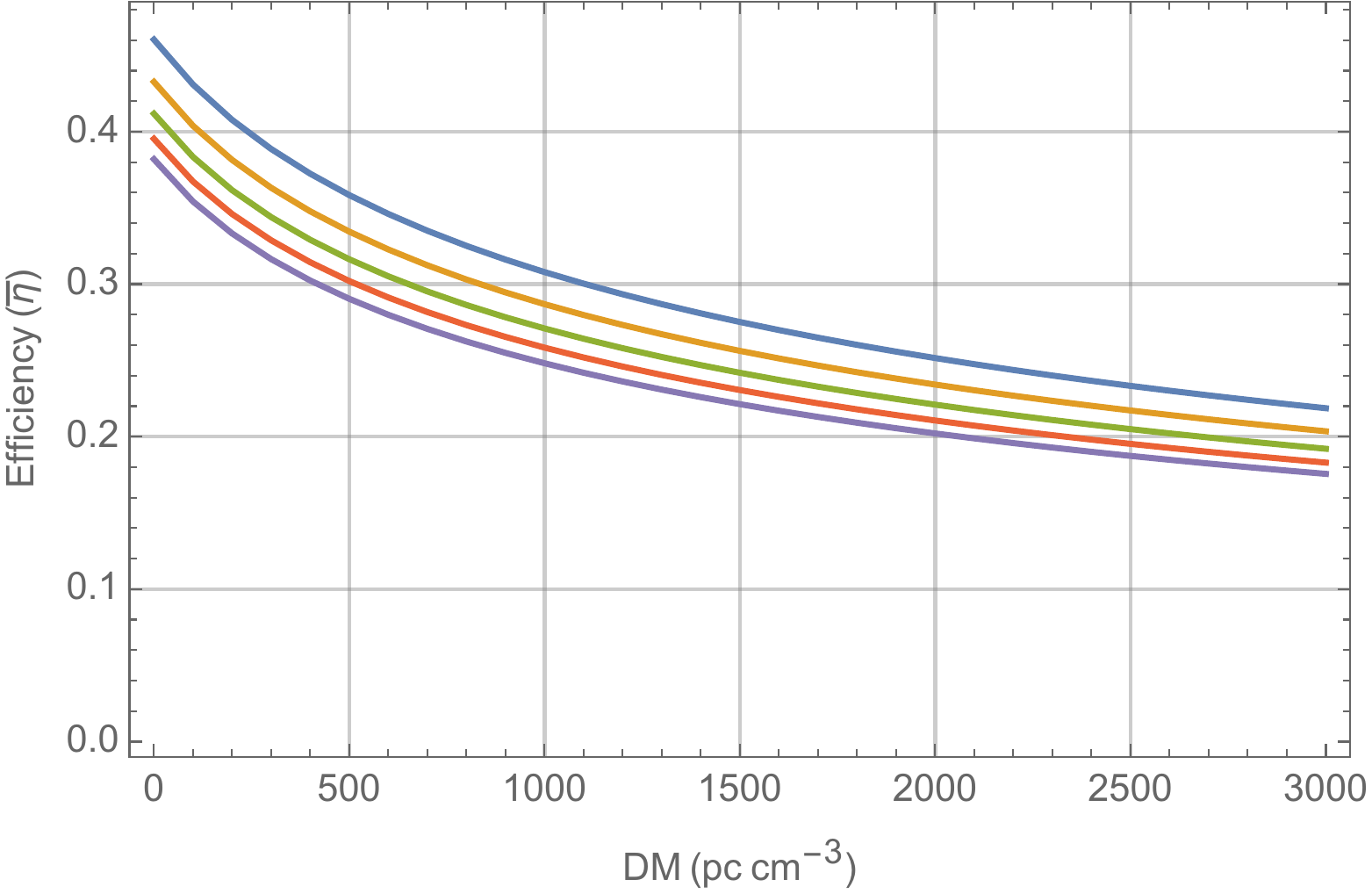}
    \caption{Representative response curves for ASKAP using the population pulse-width mean and standard deviation of $\mu = 3.44$\,ms and $\sigma=2.66$\,ms respectively. The left panel depicts the pulse-width mean changing over the range $\mu=3.34$\uc{1.0}\,ms in steps of 0.5\,ms (blue through violet in ascending order) whilst holding $\sigma$ constant at $\sigma=2.66$\,ms. The right panel depicts the response curves for a constant mean of $\mu = 3.44$\,ms whilst varying $\sigma$ over the range $\sigma=$ 2.66\uc{0.5}\,ms in steps of 0.25\,ms (blue through violet in ascending order). A similar behaviour is exhibited for the Parkes response curves hence only ASKAP is included.}
    \label{fig:CalcResponses}
\end{figure}

%==================================
\subsection{The DM Distribution Parameters}
We conduct an initial comparison of the \ME model against the Parkes and ASKAP data by exploring the semi-constrained parameter space of fluence spectral index, energy power-law index, survey fluence limit and redshift evolutionary model. The objective here being to obtain a qualitative understanding of the influence of the fit parameters on the DM distribution.

We assume a power law distribution in burst energy referenced to a $10$\funits source at $z = 1$, with a fluence spectral index of $\alpha=0$. We extend the energy curve one decade above and seven decades below the reference source, via the relation $E_{\nu} = 4 \pi D_{L}^{2}(z) F_{\nu} / (1+z)^{2-\alpha}$, corresponding to $E_{\text{min}} \approx 1.28 \times 10^{22}\,\text{J Hz}^{-1}$ and $E_{\text{max}} \approx 1.28 \times 10^{29}\,\text{J Hz}^{-1}$. We therefore span the upper region found by \citet[][Figure 2]{Shannon_etal_2018}, wherein an absence of sources above $\sim10^{27}\,\text{J Hz}^{-1}$ was noted. We find the shape of the modelled DM distribution to be insensitive to $E_{\text{min}}$ for at least seven decades below our chosen reference and that $E_{\text{max}}$ affects the distribution shape beyond approximately two decades above the upper limit found by \cite{Shannon_etal_2018}. In this latter case, the peak height becomes suppressed and the distribution tail extended at higher DMs.

The effective thresholds for Parkes and ASKAP, $F_{\text{0,P}}$ and $F_{\text{0,A}}$, are functions of the slope of the source counts distribution \citep{Macquart_Ekers_2018_1}; they vary in the ranges  $F_{\text{0,P}} = 3 \pm 2$ \funits and $F_{\text{0,A}} = 26 \pm 5$ \funits \citep{James_etal_2018}.

\begin{figure*}
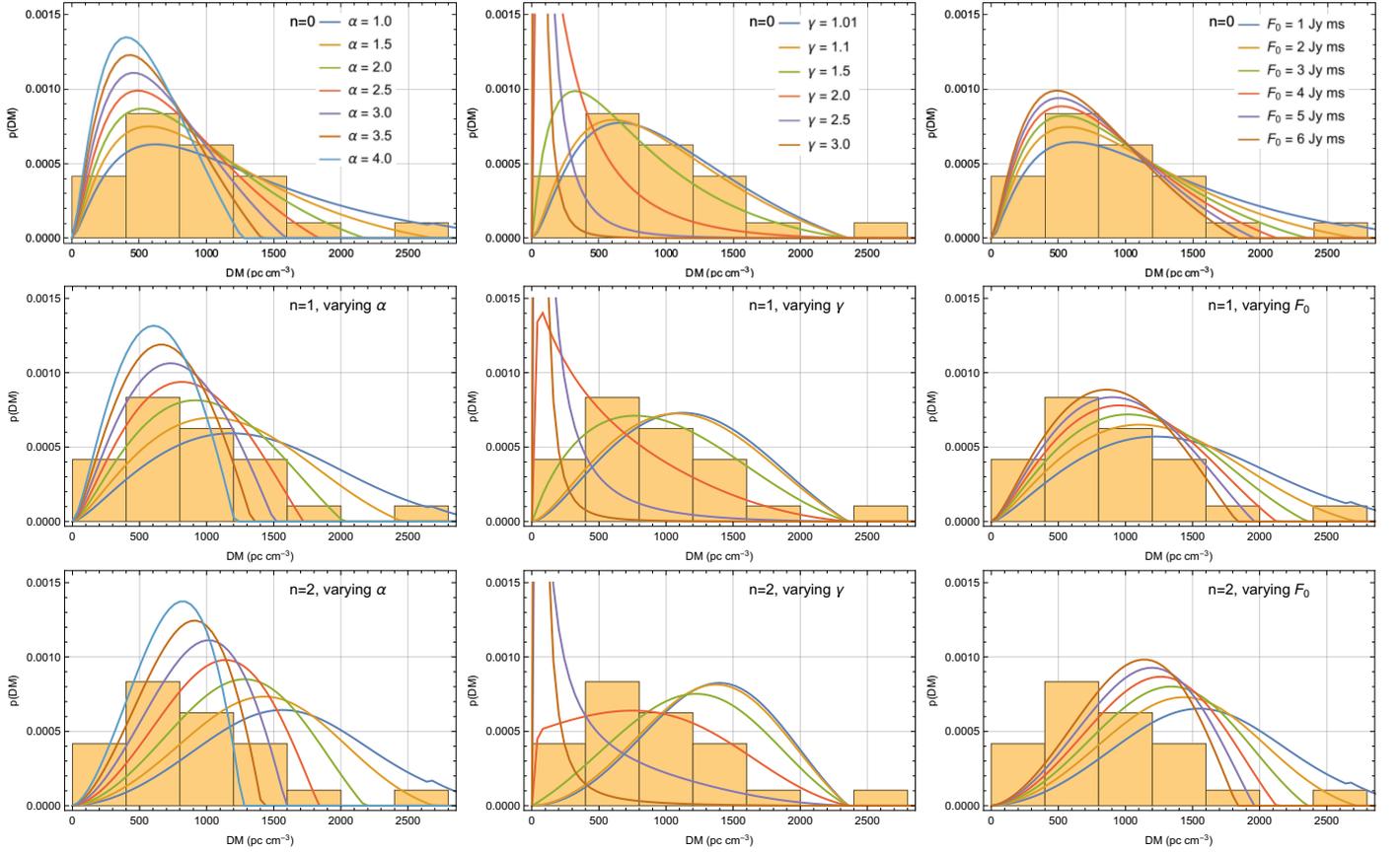

\begin{tabular}{iii}
    Parkesn0VariableAlpha-Family & Parkesn0VariableGamma-Family & Parkesn0VariableF0-Family \\
    Parkesn1VariableAlpha-Family & Parkesn1VariableGamma-Family & Parkesn1VariableF0-Family \\
    Parkesn2VariableAlpha-Family & Parkesn2VariableGamma-Family & Parkesn2VariableF0-Family \\
\end{tabular}
\caption{Figures depicting the family of curves pertaining to the modelled Parkes DM distributions and the observed histogram for various scenarios in order to explore parameter space and understand overall trends. Rows correspond to the evolutionary model scenarios $\{\psi_{n}(z) : n \in \{0, 1, 2\}\}$, respectively. Columns represent the family of curves related to changing parameters over $\alpha \in \{1.0, 1.5, 2.0, 2.5, 3.0, 3.5, 4.0\}$, $\gamma \in \{1.01, 1.1, 1.5, 2.0, 2.5, 3.0\}$ and $F_{0} \in \{1, 2, 3, 4, 5, 6\}$\funits respectively; $\alpha$, $\gamma$ and $F_{0}$ are otherwise held constant at $\alpha=1.5$, $\gamma=1.2$ and $F_{0}=3$\,Jy\,ms.}
\label{fig:ParkesFamilyScenarios}
\end{figure*}

\begin{figure*}
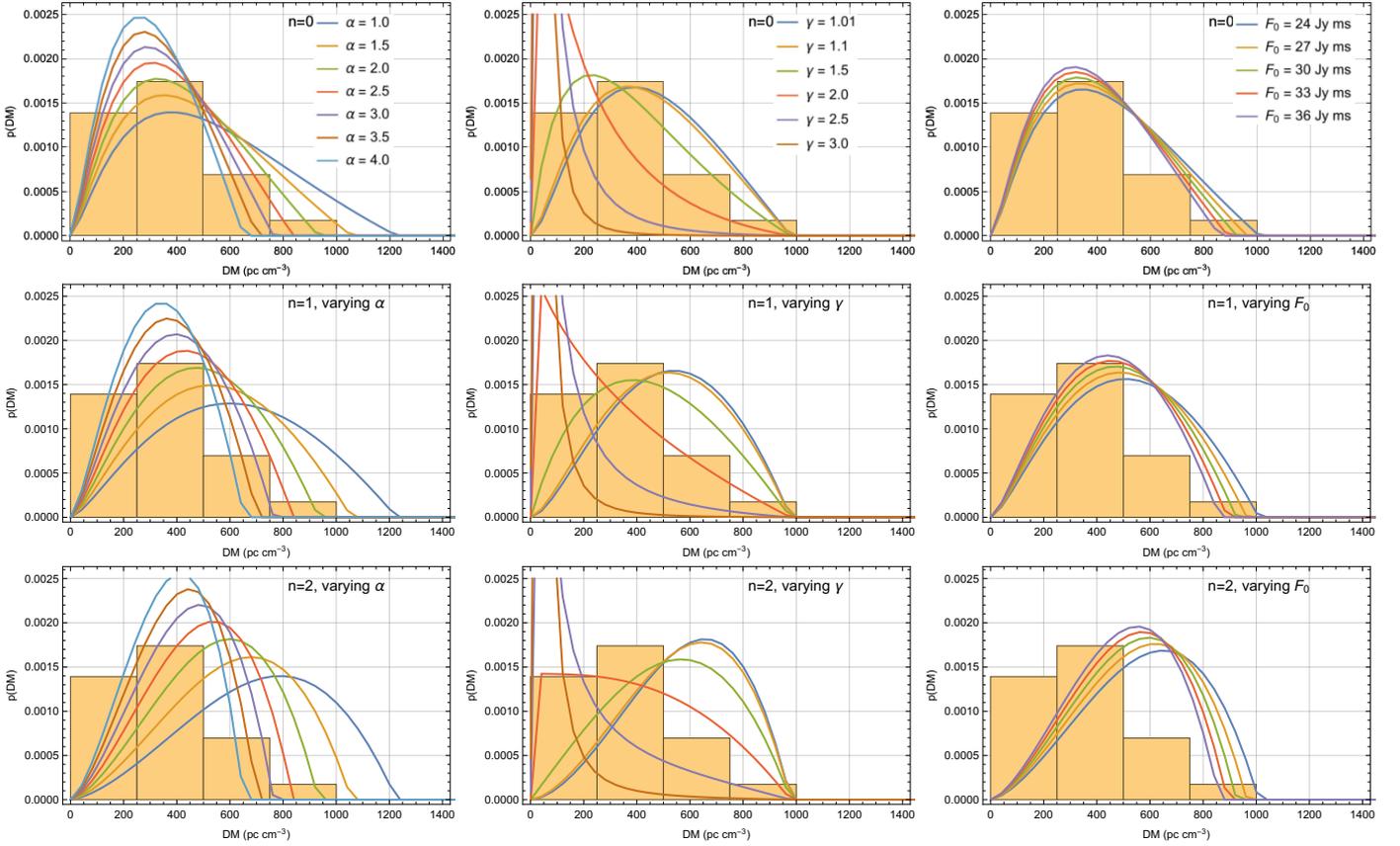

\begin{tabular}{iii}
    ASKAPn0VariableAlpha-Family & ASKAPn0VariableGamma-Family & ASKAPn0VariableF0-Family \\
    ASKAPn1VariableAlpha-Family & ASKAPn1VariableGamma-Family & ASKAPn1VariableF0-Family \\
    ASKAPn2VariableAlpha-Family & ASKAPn2VariableGamma-Family & ASKAPn2VariableF0-Family \\
\end{tabular}
\caption{Figures depicting the family of curves for the modelled ASKAP DM distributions and the observed histogram for various scenarios in order to explore parameter space and overall trends. Rows correspond to the evolutionary model $\{\psi_{n}(z) : n \in \{0, 1, 2\}\}$, respectively. Columns represent changes to the family of curves pertaining to changing parameters over the ranges $\alpha \in \{1.0, 1.5, 2.0, 2.5, 3.0, 3.5, 4.0\}$, $\gamma \in \{1.01, 1.1, 1.5, 2.0, 2.5, 3.0\}$ and $F_{0} \in \{24, 27, 30, 33, 36\}$\funits respectively; $\alpha$, $\gamma$ and $F_{0}$ were otherwise held constant at $\alpha=1.5$, $\gamma=1.2$ and $F_{0}=26$\,Jy\,ms.}
\label{fig:ASKAPFamilyScenarios}
\end{figure*}

We account for the effects of finite instrumental resolution by multiplying the modelled (intrinsic) FRB DM distributions by their corresponding instrument response -- i.e., via $\overline{\eta}(DM) \cdot \: dR_{F}/dDM$ (see eq.(\ref{eqn:DMEfficiency}), eq.(\ref{eqn:DMDistribution}) and Table \ref{tab:SensitivityFit}) -- before comparing the modelled DM distributions with the observed histograms. Figures \ref{fig:ParkesFamilyScenarios} and \ref{fig:ASKAPFamilyScenarios} depict the family of curves generated for the Parkes and ASKAP events respectively. In each of the array of figures, rows correspond to the redshift evolutionary models, $\{\psi_{n}(z) : n \in \{0, 1, 2\}\}$ respectively, whilst columns pertain to changing $\alpha$ and $\gamma$ respectively.

We make the following general observations regarding these scenarios: a change in the spectral index has a significant effect on the lateral displacement of the DM distributions. For the chosen burst energy distribution, a spectral index of $\alpha \sim \left[1.5,2.0\right]$ aligns the peak of the distributions to the observed DM histograms, a result in agreement with $\alpha=1.5^{+0.3}_{-0.2}$ found by \citet{Macquart_etal_2019}. The evolutionary model, acting via the redshift-dependent terms of eq.(\ref{eqn:RedshiftDistribution}), also has a significant effect on lateral displacement, however, it has the additional effect of skewing the DM distributions to higher DM as the evolutionary model transitions from $\psi_{0}(z)$ through to $\psi_{2}(z)$. Accordingly, the shape of the underlying DM distribution significantly deviates from that of the observed histogram -- an effect that becomes more pronounced in the higher fluence survey limit regime relevant to ASKAP. The scenarios of no redshift evolution ($n=0$) or linear redshift evolution ($n=1$) with respect to the CSFR tends to yield a closer overall fit in terms of peak alignment and distribution shape (i.e., the lack of cuspiness) with respect to the observed histograms, particularly as $\gamma \rightarrow 2.5$ -- a trend seen across both survey-limit regimes.

%==================================
\subsection{Parameter Fitting}
With this qualitative insight, we fit the modelled DM distributions to the observed histograms to determine the best-fit parameters of $\hat{\alpha}$ and $\hat{\gamma}$ for each redshift evolutionary model as described below.

From our exploration of parameter space, we note that Figures \ref{fig:ParkesFamilyScenarios} and \ref{fig:ASKAPFamilyScenarios} indicate the estimated DM distributions vary slowly with $F_{\text{0,P}}$ and $F_{\text{0,A}}$ and that the estimated likelihoods are insensitive to these parameters. We therefore fix $F_{\text{0,P}} = 3$ \funits and $F_{\text{0,A}} = 26$ \funits throughout. We fit both FRB datasets simultaneously, on the assumption that the energy power-law index and fluence spectral index are common to the FRB population. We compute the p-value (representing likelihood) that the observed data is drawn from the distribution predicted by a given model, with parameters drawn over the semi-constrained parameter-space grid $\{(\alpha, \gamma, p)_{i}, \forall i\}$ at a grid resolution of $\Delta \alpha = \Delta \gamma = 0.02$.

We initially compare four fitting methods: three bin-independent methods, viz., Kolmogorov-Smirnov (K-S), Anderson Darling and Watson U Square and the bin-dependent Pearson $\chi^{2}$ method. This was undertaken to ensure robustness of the fits given the relatively low number of samples in the dataset. We determine the best fit parameters via the product of the p-values for each fit. During the fitting process we correct the intrinsic DM distribution for telescope sensitivity and re-normalise; the purpose being to match the shape of the distributions, since the absolute FRB event rates are difficult to calibrate \citep[see, e.g.,][]{James_etal_2018}.

We assess fitting robustness in two primary ways: first, we check that the fitting performance is robust to pulse-width variation, by utilising a mean pulse-width one decade lower than that determined for the FRB dataset\footnote{The effect of pulse-widths below the instrument temporal resolution is of primary interest here, since the distribution is poorly characterised observationally on short timescales.}, viz., $\mu=0.34$\,ms. We find the results to be stable to this effect: the telescope sensitivity curves retain their overall shape and the introduced offset is negated during normalisation. That is, the DM probability distribution is sensitive to shape, not to the differences in the absolute burst detection rates between the Parkes and ASKAP samples. Second, we compare the results of the four fitting methods. All four methods yield broadly comparable results however we note that the Pearson $\chi^{2}$ method is affected by the choice of bin sizes, due to the low number of FRBs in the sample set, causing p-values to fluctuate. We select the K-S method throughout and recommend bin-independent methods be considered in situations where the sample size may be small or otherwise sensitive to the choice of data binning.
% it is therefore discarded. Of the remaining three bin-independent tests, the Watson U Square test is used primarily for circular data (e.g. as a function of angle), and is neither applicable nor sensitive in this case. Comparing the Anderson Darling and K-S tests, the former is more sensitive to the tail of the distribution. While our assumption of the DM-z relation likely holds for the majority of the sample, it is possible for a large excess DM to result in an outlier. To avoid such a case potentially biasing out results, we select the (bin-independent) K-S method. We recommend that bin-independent methods be considered in situations where the sample size may be small or otherwise sensitive to the choice of data binning.

\begin{figure}
    \centering
    \includegraphics[scale=\DefaultPlotSize]{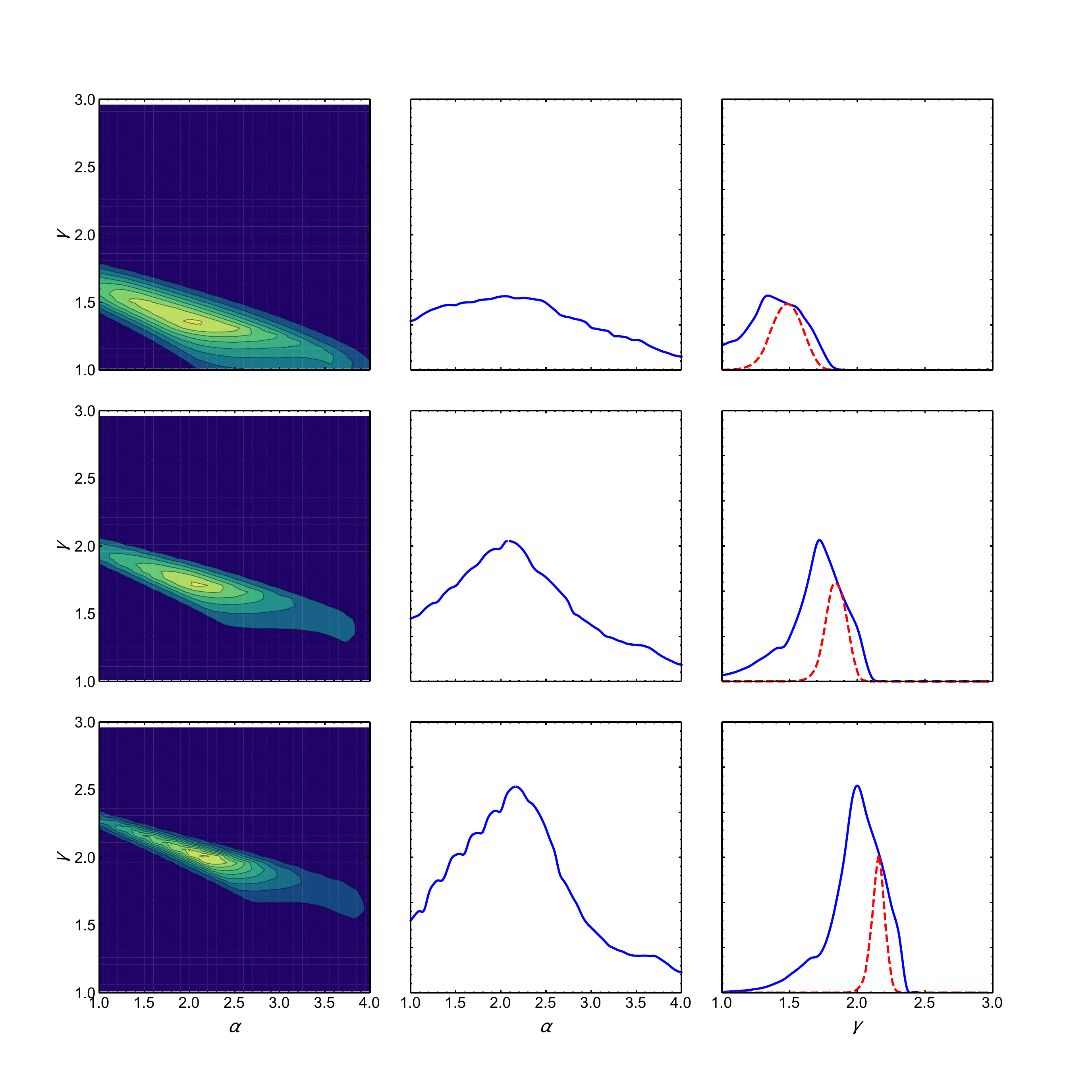}
    \caption{Confidence regions for the K-S test-based fits, along with the supremum of the joint p-values (likelihoods) for the simultaneously estimated parameters $\alpha$ and $\gamma$, for the redshift evolutionary models $n \in \{0, 1, 2\}$. Rows correspond to the models $n=0$ (top), $n=1$ (middle) and $n=2$ (bottom) while columns represent the joint p-value contour plots for $\alpha$ vs. $\gamma$ (left) and the supremum of those p-values for $\alpha$ (centre) and $\gamma$ (right). The dashed red curve in the right column represents the likelihood distribution of the constraint $p(\gamma|\alpha=1.5)$. The survey fluence limits for the Parkes and ASKAP telescopes are held fixed at $F_{0,\text{P}} = 3$\funits and $F_{0,\text{A}} = 26$\funits during the fitting process.}
    \label{fig:ErrorPanelFit}
\end{figure}

Initially, we search the ranges $\gamma \in \left[1.01, 3.0\right]$ and $\alpha \in \left[1.0, 4.0\right]$ to ensure the parameter searches are not overly constrained, and to avoid omitting the best global fit or biasing the parameter estimates. (The lower bound of $\gamma=1.01$ was chosen to avoid the pole at $\gamma=1$ of eq.(\ref{eqn:RedshiftDistribution}).) Table \ref{tab:FinalParams} summarises the overall best-fit parameters attained, along with their 68\% confidence intervals using the K-S test. The corresponding confidence regions for the redshift evolutionary models $n \in \{0, 1, 2\}$ are depicted in Figure \ref{fig:ErrorPanelFit}. We subsequently further constrain $\alpha=1.5$, consistent with \citet{Macquart_etal_2019}, and recompute $\hat{\gamma}$ for the same models, which are also given in the second half of Table \ref{tab:FinalParams}.

\begin{figure*}
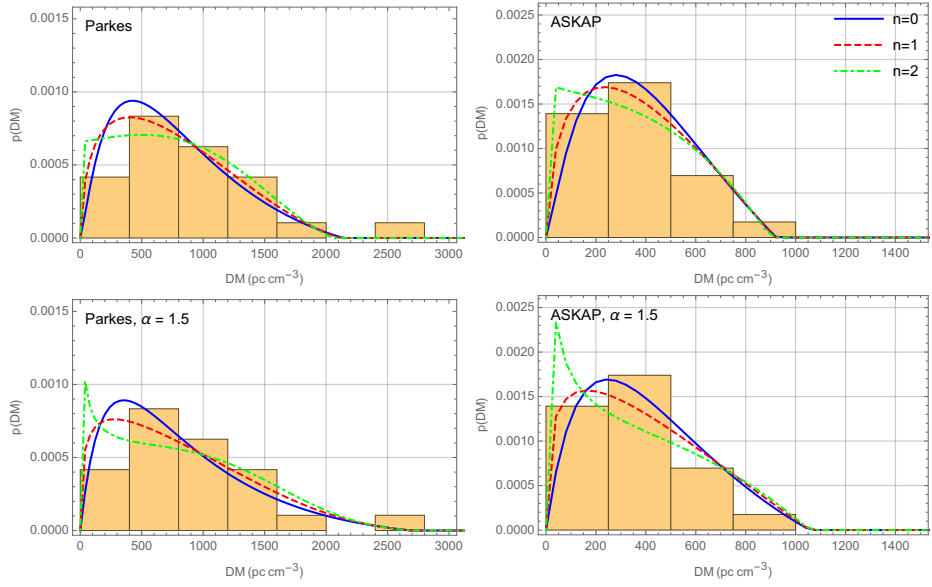

\begin{tabular}{iii}
    Unconstrained-KS-Parkes & Unconstrained-KS-ASKAP \\
    Constrained-KS-Parkes   & Constrained-KS-ASKAP   \\
\end{tabular}
\caption{The Parkes (left panels) and ASKAP (right panels) modelled DM distributions and corresponding observed histograms using the best-fit parameters determined from the K-S test-based fits for each of the redshift evolutionary models. Solid blue curves represent no redshift evolution ($n=0$), dashed red curves linear redshift evolution ($n=1$) and green dot-dashed curves quadratic redshift evolution ($n=2$) with respect to the CSFR. The top two figures relate to the best-fit parameters attained from the full search range and correspond to $\{(n, \alpha, \gamma)\} \in \{(0, 2.0,1.3),(1, 2.1,1.7),(2, 2.2,2.0)\}$ whilst those in the bottom panels relate to further constraining $\alpha=1.5$, viz., $\{(n, \alpha, \gamma)\} \in \{(0, 1.5,1.5),(1, 1.5,1.8),(2, 1.5,2.2)\}$ -- see Table \ref{tab:FinalParams}.}
\label{fig:ParkesASKAPFinalFig}
\end{figure*}

The best-fit DM distributions and observed histograms pertaining to the parameters listed in Table \ref{tab:FinalParams} are depicted in Figure \ref{fig:ParkesASKAPFinalFig} for both the Parkes (left panels) and ASKAP (right panels) telescopes. Plots in the top panels pertain to fits relating to the broader parameter space whilst those in bottom panels relate to further constraining $\alpha=1.5$.

\begin{table*}
\caption{The K-S test-based best-fit parameters and their 68\% confidence intervals for the Parkes and ASKAP FRB events, for different redshift evolutionary models. We simultaneously fit the FRB sample sets based on the assumption that the FRB events are drawn from the same population. The best-fit parameters relate to: (i) the semi-constrained parameter space of $\gamma \in \left[1.01, 3.0\right]$ and $\alpha \in \left[1.0, 4.0\right]$; and (ii) applying a further constraint of $\alpha=1.5$, consistent with \citet{Macquart_etal_2019}.}
\renewcommand{\arraystretch}{1.5}
\label{tab:FinalParams}
\begin{threeparttable}[b]
\begin{tabular}{ccccl}
\midrule
$n$ & p-value & $\hat{\alpha}$ & $\hat{\gamma}$ & Evolution\tnote{$\dagger$}\\
\midrule
0 & 0.164 & 2.0\uclu{1.0}{1.5} & 1.3\uclu{0.2}{0.3} & None      \\
1 & 0.311 & 2.1\uclu{1.1}{0.9} & 1.7\uclu{0.2}{0.2} & Linear    \\
2 & 0.457 & 2.2\uclu{1.0}{0.7} & 2.0\uclu{0.1}{0.3} & Quadratic \\
\midrule
0 & 0.146 & 1.5\tnote{$\ddagger$} & 1.5\uclu{0.2}{0.2} & None      \\
1 & 0.218 & 1.5\tnote{$\ddagger$} & 1.8\uclu{0.1}{0.1} & Linear    \\
2 & 0.305 & 1.5\tnote{$\ddagger$} & 2.2\uclu{0.1}{0.1} & Quadratic \\
\midrule
\end{tabular}
\begin{tablenotes}
\item [$\dagger$] Redshift evolution re CSFR (see eq.(\ref{eqn:CSFR})).
\item [$\ddagger$] A set constraint.
\end{tablenotes}
\end{threeparttable}
\end{table*}

%==================================
%\pagebreak
\section{Discussion}
Whilst the DM-redshift relation is generally not expected to be bijective, except when averaged over many lines of sight \citep[see e.g.][]{McQuinn_2014}, we nonetheless make this assumption advisedly on the basis that the DM dataset utilised exhibits high DMs with low DM dispersion and on the basis of recent work establishing a DM-redshift relation for localised FRBs \citep{Macquart_etal_2020}. We further assume an homogeneous IGM and determine the best-fit population parameters of fluence spectral index, $\hat{\alpha}$, and energy power-law index, $\hat{\gamma}$, for an assumed energy curve. We determine these parameters simultaneously using the joint p-value via the K-S test and for different redshift evolutionary models, by comparing the \ME modelled DM distribution shapes with the observed histograms.

The approach adopted circumvents a number of key unknowns regarding the FRB population. First, the model allows for the ready generation of the DM distribution using few population and instrument parameters (viz., $\alpha$, $\gamma$ \& $F_{0}$) with a relatively simple assumption for the energy curve cut-offs, $E_{\text{min}}$ and $E_{\text{max}}$, even though their values are not well established. Second, it permits direct comparison between datasets from telescopes of different survey sensitivities, obviating the need to address difficulties around calibration: the absolute FRB event rates are not required. By using the relative FRB event rates, and given the demonstrated robustness to pulse-widths smaller than the instrument resolution, we find the overall sensitivity curve shapes do not change significantly (i.e., shape changes are higher-order effects) and they are subsequently normalised out during the fitting process -- a process insensitive to the specific (log normal) distribution chosen. We find simultaneously fitting the FRB data for both telescopes, using the K-S method, to be robust. Third, even though the \ME model derives $\alpha$ principally on cosmological k-correction grounds (i.e., it is measured via the correction $(1+z) L_{(1+z)_{\nu}}/L_{\nu}$ made because the radiation is observed in a different band from that emitted by the source), even with an assumed energy curve, it compares favourably to values determined from independent means -- e.g., $\alpha=1.5_{+0.3}^{-0.2}$ \citep{Macquart_etal_2019} and $\alpha=1.8 \pm 0.3$ \citep{Shannon_etal_2018}. Furthermore, the fits suggest that the distribution of burst energies for the population is relatively flat (viz., $\gamma < 2$) in this DM regime.

Other authors also fit the DM distribution for Parkes and ASKAP data. \citet{Lu_Piro_2019} examines the DM distribution of ASKAP samples (only), finding $\gamma=1.6 \pm 0.3$ with a fitted $\log_{10} E_{\rm max}=27.1_{0.7}^{+1.1}$\,J\,Hz$^{-1}$ (68\% confidence) -- two orders of magnitude less than that used herein. Their model is broadly similar to the \ME model, with the following key differences: (a) the width distribution of the FRBs, hence its effect on sensitivity, is not included; (b) an exponential tail to the luminosity function beyond $E_{\rm max}$ is used, rather than a sharp cut-off as used herein; (c) the authors further assume, but do not fit, $\alpha=1.5$; and (d) the authors study source evolution via $\psi(z)\sim(1+z)^\beta$, finding $\beta=0.8^{+2.6}_{-2.9}$ (approximately corresponding to $n=0.3^{+1.0}_{-1.1}$). Nonetheless, those results, together with the large error of their fits, are comparable to ours. In \citet{Luo_etal_2020}, the authors consider a wider sample of FRBs, including ASKAP and Parkes observations, as well as those from several other instruments. A key difference of that treatment is the inclusion of DM scatter about the expectation for a given redshift, however they assume a flat spectrum ($\alpha=0$) and do not consider source evolution. These authors find $E_{\rm max}=2.9^{+11.9}_{-1.7} \times 10^{28}$\,J\,Hz$^{-1}$ (converted assuming a 1\,GHz bandwidth), and $\gamma=1.79^{+0.35}_{-0.31}$ -- results also comparable with those found here.

Given the results attained, our assumption that the telescopes observe the same FRB population is consistent with the observations. As discussed at length in \citet{Macquart_Ekers_2018_2}, the behaviour of the DM distribution depends upon the slope of the FRB energy (or luminosity) function. Between $\gamma \approx 2-3$ there is expected to be a dramatic change in character of the DM distribution, with flatter distributions probing to higher redshifts, where they contain a large fraction of observed events at large distances. At a critical value of $\gamma=2.5$ there is no distance dependence on fluence, hence no information on evolution. For the high redshift evolution fits (viz., $n=2$), the solutions push $\gamma$ closer to this critical value, and whilst this may be the correct interpretation, it is more likely to be finding a solution that is independent of the imposed evolution.

From Figure \ref{fig:ParkesASKAPFinalFig}, it can be seen that this effect predicts an excess of FRBs in the nearby Universe ($DM \sim 0$), particularly for the ASKAP sample. This effect does not appear to be observed: recent localisations of FRBs by  ASKAP \citep{Bannister_etal_2019,Prochaska_etal_2019,Macquart_etal_2020} do not show this excess. Whilst we therefore cannot definitively exclude a strongly evolving population (viz., $n=2$), it does seem unlikely. Future surveys with a greater fluence range will reduce this degeneracy. Conversely, for steeper distributions, observations with higher sensitivity will be dominated by nearby events. In our analysis, we determine the FRB energy function to be relatively flat (see Table \ref{tab:FinalParams}), suggesting FRBs should be readily detectable to higher redshifts. Accordingly, Parkes and other more sensitive telescopes such as CHIME and FAST may be better able to discriminate the effects of population evolution, thereby aiding in the selection of progenitor model classes as the FRB event dataset grows.

We favour the case of no redshift evolution (i.e., $n=0$), based predominantly on constraining $\alpha=1.5$ -- see Table \ref{tab:FinalParams}, Figure \ref{fig:ParkesASKAPFinalFig} and the likelihood curves for $p(\gamma|\alpha=1.5)$ of Figure \ref{fig:ErrorPanelFit}. Despite the relative p-values between redshift evolutionary models indicated in Table \ref{tab:FinalParams}, we disfavour the model of quadratic redshift evolution ($n=2$) with respect to the CSFR due to the cuspiness exhibited in the modelled DM distribution (see Figure \ref{fig:ParkesASKAPFinalFig}, green dot-dashed curves). The cuspiness being a direct consequence of $\gamma \ge 2$, representing an aggregation of FRBs at low DM, which is not present in the observed histograms. We estimate the mean redshift probed by the Parkes and ASKAP telescopes to be approximately $0.62$ and $0.32$, respectively.

\begin{figure}
    \centering
    \includegraphics[scale=\BigPlotSize]{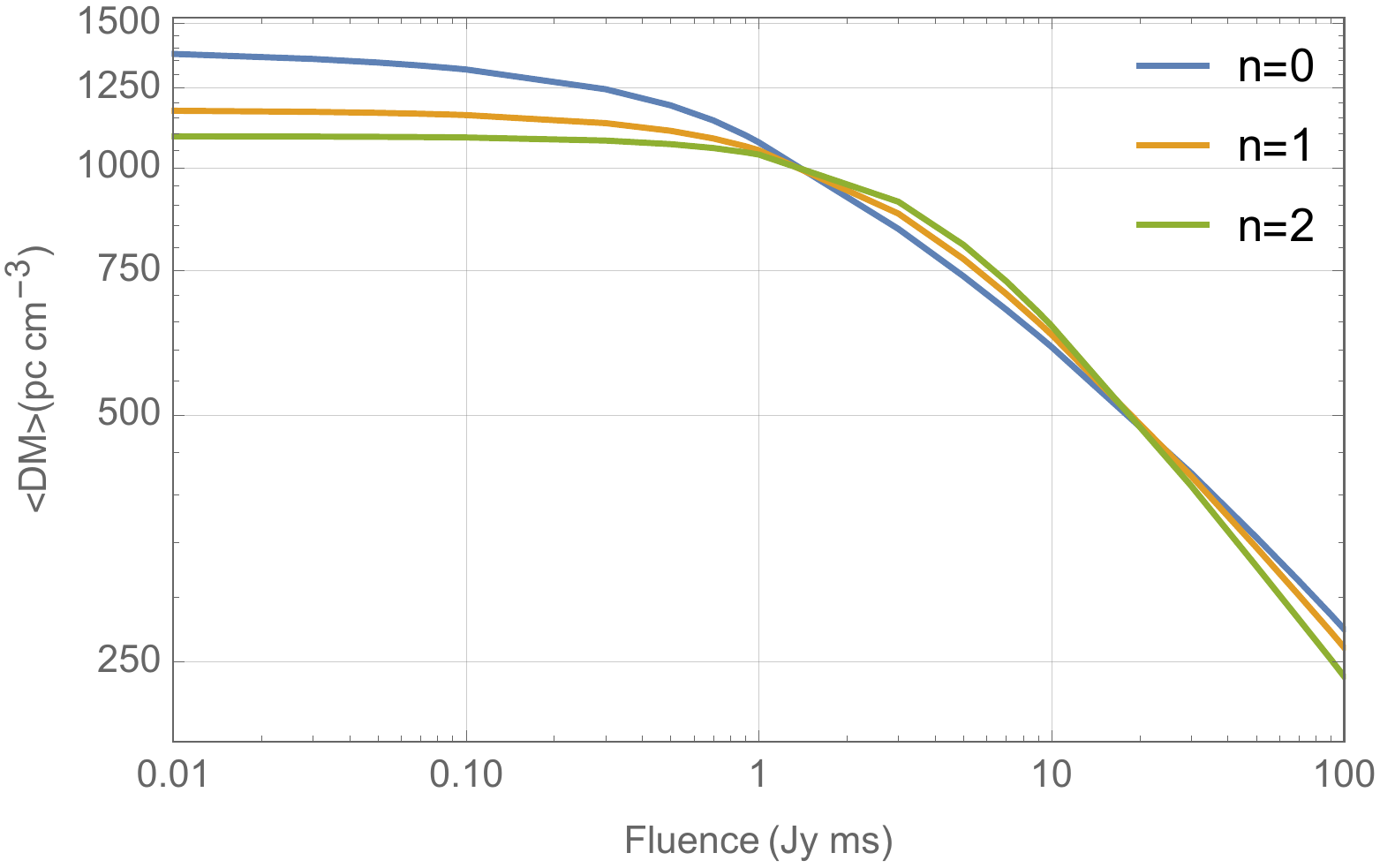}
    \caption{The expectation of the DM distribution, $\langle DM \rangle$, as a function of survey fluence limit, $F_{0}$, for an homogeneous IGM. We use the best-fit energy power-law index, $\hat{\gamma}$, for the constrained spectral index of $\alpha=1.5$ and for the three redshift evolutionary models listed in Table \ref{tab:FinalParams} via eq.(\ref{eqn:DMExpectation}). The blue curve represents no redshift evolution ($n=0$), orange linear redshift evolution ($n=1$) and green quadratic redshift evolution ($n=2$) with respect to the CSFR. We extend the fluence survey limit down to anticipated SKA levels for reference purposes. Here, we use the Parkes response curve, as the corresponding ASKAP response-based curves are ostensibly the same, hence are omitted.}
    \label{fig:FluenceDMCurve}
\end{figure}

Motivated by the results of the fits, we further determine the DM expectation, $\langle DM \rangle$, that a survey-limited telescope is expected to probe using eq.(\ref{eqn:DMExpectation}). We compute $\langle DM \rangle$ for fluence survey limits extending down to the anticipated regime of the Square Kilometre Array (SKA), viz., $F_{0} \sim 0.01$ Jy ms, as shown in Figure \ref{fig:FluenceDMCurve}. Here, we use the best-fit energy power-law index, $\hat{\gamma}$, for the constrained fluence spectral index of $\alpha=1.5$, for the redshift evolutionary models of Table \ref{tab:FinalParams}, and use the Parkes response curve (the ASKAP response-based curves, yielding ostensibly the same result, are omitted).

\begin{equation}
    \langle DM(F_{\nu}>F_{0})\rangle = \int_{0}^{\infty} DM' \, \dfrac{dR_F}{dDM'}(DM'; \alpha, \gamma,n, F_{0}, F_{\text{min}}, F_{{max}})\,dDM'.
\label{eqn:DMExpectation}
\end{equation}
\\

%==================================
% \pagebreak
\section{Conclusion}
We compare the observed DM histograms of two FRB sample sets detected by the ASKAP and Parkes radio telescopes with distributions of the \ME model and exploit the fact that the telescopes have different survey fluence limits.

After accounting for temporal and spectral resolution of the data, we show that the modelled distributions fit the observed histograms well, and that by comparing the distribution shapes, the absolute FRB event rate is not required -- providing a significant advantage in not having to address calibration complexities or unknown survey rate corrections. In this DM regime, DM does seem to be a reasonable proxy for redshift thereby providing additional evidence over direct measurements for a handful of localised FRBs \citep[cf.][]{Macquart_etal_2020}, that the IGM does indeed dominate the DM budget for the FRB population as a whole.

After fitting the modelled distributions to the observed datasets simultaneously, we determine the best-fit population parameters of fluence spectral index and energy power-law index, for an assumed energy curve, and for different redshift evolutionary models.

The fluence spectral index, manifest as a k-correction in our analysis, models the value obtained independently by direct fits to burst spectra. This seems remarkable given the irregular burst spectra often measured and that it approximates values determined by observationally independent means \citep[see e.g.,][]{Macquart_etal_2019,Shannon_etal_2018}. Based on these results, we find that the two telescopes likely do observe the same FRB population and that the energy curve may indeed be relatively flat in this DM regime.

Fits for an FRB population evolving faster than the star-formation rate predict $\gamma\approx2.2$, leading to an expectation of many FRBs occurring in the nearby Universe, contrary to observations. After constraining the fluence spectral index to $\alpha=1.5$, we find no evidence that the FRB population evolves faster than linearly with respect to the star formation rate, which places further constraints on progenitor classes.

Motivated by the performance of the \ME model, and the prospect of much larger FRB sample-sets in future, it seems worthwhile for future studies to investigate more realistic FRB evolutionary scenarios, such as those in which FRBs may exhibit a substantial finite time to evolve from the epoch at which their progenitors form. A further analysis of the FRB population, using telescopes of significantly different sensitivities, such as with FAST and CHIME would also be worthwhile: more sensitive telescopes will be more effective in discriminating the effect of changing  sensitivity, $F_{0}$, as can be seen in Figure \ref{fig:ParkesFamilyScenarios} (bottom right). Other areas of further investigation would include: (i) exploring the effect of $E_\text{max}$ of the burst energy distribution; (ii) examine the effects of the degeneracy in $\gamma$ and its mitigation; and (iii) extending the analysis to incorporate the effect that IGM inhomogeneities may have on the DM distribution \citep[see \textsection4 of][]{Macquart_Ekers_2018_2}.

\clearpage
%==================================
%\clearpage
\section*{Acknowledgements}
This research was partly supported by the Australian Research Council through grant DP180100857. WA acknowledges the contribution of an Australian Government Research Training Program Scholarship in support of this research. \textit{In Memoriam}: Vale J-PM, our dear friend and esteemed co-author. Like the FRBs you studied, you leave a blazing signal for others to follow; your enthusiasm and insight profound.

%==================================
\section*{Data availability}
Data underlying this article are available in the article.



\bibliographystyle{mnras}
\bibliography{References.bib} % if your bibtex file is called example.bib

% \begin{thebibliography}{99}
% \bibitem[\protect\citeauthoryear{Author}{2012}]{Author2012}
% Author A.~N., 2013, Journal of Improbable Astronomy, 1, 1
% \bibitem[\protect\citeauthoryear{Others}{2013}]{Others2013}
% Others S., 2012, Journal of Interesting Stuff, 17, 198
% \end{thebibliography}

%%%%%%%%%%%%%%%%%%%%%%%%%%%%%%%%%%%%%%%%%%%%%%%%%%

% If you want to present additional material which would interrupt the flow of the main paper,
% it can be placed in an Appendix which appears after the list of references.

%%%%%%%%%%%%%%%%% APPENDICES %%%%%%%%%%%%%%%%%%%%%
\clearpage
% \pagebreak
\appendix
\section{DM Response Parameters}
Input and fitted parameters attained by injecting artificial bursts with a synthetic flat time-frequency profile at random times into the time-frequency dynamic spectrum and an incoherent de-dispersion performed for both the Parkes and ASKAP telescopes.

\begin{table}
\caption{Input and best-fit DM response parameters for Parkes and ASKAP telescopes, apropos eq.(\ref{eqn:DMEfficiency}).}
\begin{tabular}{llll}
\hline
Parameter & Parkes & ASKAP & Units\\
\hline
\underline{\textit{Instrumental}}: &&& \\
$\nu_{r}$  & 0.39   & 1        & MHz \\
$\nu_{c}$  & 1.361  & 1.283    & MHz \\
$t_{r}$    & 0.064  & 1.266    & ms  \\
\underline{\textit{Fitted}}:  &&&    \\
$\eta_{0}$ & 0.72 & 0.76   &     \\
$c_{1}$    & 0.94 & 0.94   &     \\
$c_{2}$    & 0.05 & 0.37   &     \\
\hline
\end{tabular}
\label{tab:SensitivityFit}
\end{table}

\clearpage
% \pagebreak
% Don't change these lines
\bsp    % typesetting comment
\label{lastpage}
\end{document}